\documentclass[journal]{IEEEtran}
\usepackage{cite}
\usepackage{amsmath,amssymb,amsfonts}
\usepackage{hhline}
\usepackage{tabularray}
\usepackage{algorithmic}
\usepackage{graphicx}
\usepackage{optidef}
\usepackage{enumitem} 
\usepackage{float}
\usepackage{svg}  
\usepackage{float} 
\usepackage{placeins} 
\usepackage[ruled,vlined]{algorithm2e}
\usepackage{amsmath}

\usepackage{multirow}
\usepackage{textcomp}
\usepackage{xcolor}
\usepackage{subcaption}
\usepackage{bm}
\def\BibTeX{{\rm B\kern-.05em{\sc i\kern-.025em b}\kern-.08em
    T\kern-.1667em\lower.7ex\hbox{E}\kern-.125emX}}
\usepackage{ragged2e}
\usepackage{booktabs}  
\usepackage{tabularx}
\usepackage{array}         
\usepackage{longtable}     
\usepackage{caption}       
\usepackage{graphicx}

\usepackage{tabularx}
\usepackage[table]{xcolor}
\usepackage{colortbl} 
\usepackage{tabularray} 
\usepackage{hhline} 
\usepackage{booktabs}

\newcolumntype{C}[1]{>{\centering\arraybackslash}p{#1}}
\newcolumntype{Y}{>{\RaggedRight\arraybackslash}X}

\usepackage{xcolor}
\usepackage{tabularx}  
\usepackage{array}
\usepackage{makecell}
\usepackage{booktabs} 

\newcolumntype{C}[1]{>{\centering\arraybackslash}m{#1}}

\newcolumntype{Z}{>{\centering\arraybackslash\hspace{0pt}}X}

\newcolumntype{L}{Z} 
\newcolumntype{M}{>{\hsize=1.00\hsize}Z} 
\newcolumntype{A}{>{\hsize=1.00\hsize}Z} 
\newcolumntype{D}{>{\hsize=1.00\hsize}Z} 

\definecolor{RowBlue}{RGB}{234,242,251}    
\definecolor{RowPeach}{RGB}{248,238,232}   
\definecolor{RowLav}{RGB}{241,237,247}     
\definecolor{RowMint}{RGB}{237,244,238}    
\definecolor{RowPink}{RGB}{244,236,239}    
\newcolumntype{C}[1]{>{\centering\arraybackslash}m{#1}}
\newcolumntype{L}{>{\raggedright\arraybackslash}X} 
\newcolumntype{Z}{>{\centering\arraybackslash}X}
\definecolor{ColAdv}{RGB}{255, 250, 205}  
\definecolor{ColLim}{RGB}{224, 255, 255}  

\ifCLASSINFOpdf
\else
\fi
\begin{document}
%
\title{Agentic AI Enabling Autonomous, Self-Organizing, and Evolving UAV Networks}
%
%
%
\author{Zhaoyang Li, Xingzhi Jin, Zijiu Yang, Qianqian Yang and Zhiguo Shi
\IEEEcompsocitemizethanks{
\IEEEcompsocthanksitem Zhaoyang Li, Xingzhi Jin, Zijiu Yang, Qianqian Yang and Zhiguo Shi are with
the College of Information Science and Electronic Engineering,
Zhejiang University, Hangzhou, China. (e-mails: \{zhaoyangli, 22560222, zijiu\_yang, qianqianyang20, shizg\}@zju.edu.cn). (Corresponding author: Qianqian Yang.)


}
}

\maketitle

\begin{abstract}
As low-altitude applications expand across emergency response, intelligent transportation, and autonomous operations, they demand communication networks that can deliver flexible, resilient, and rapidly deployable connectivity. Heterogeneous UAV networks are a promising solution, as they can dynamically provide sensing, access, relay, and backhaul functions. Yet, most existing approaches assume predefined missions, prior knowledge of user distributions, and manually configured infrastructure, making them ill suited to dynamic and initially unknown environments.
Addressing this limitation requires a shift from mission-oriented UAV deployment to autonomous network formation, in which UAVs continuously perceive their surroundings, infer evolving service demands, and self-organize network resources. Agentic AI, empowered by large language models (LLMs), offers a new foundation for this shift by integrating closed-loop perception, reasoning, planning, and execution across heterogeneous information sources. Unlike conventional optimization and learning methods designed for individual networking tasks, agentic AI can coordinate these capabilities to support sustained, network-level autonomy.
In this article, we explore agentic AI for autonomous and self-organizing heterogeneous UAV networks in low-altitude environments. Our key contribution is an LLM-assisted architecture in which a base-station-hosted agent conducts global network reasoning and autonomously reconfigures access and backhaul infrastructure. The proposed system explores unknown environments, discovers users, and deploys UAVs on demand to provide access and establish end-to-end backhaul connectivity. A case study illustrates how this agentic-AI-driven approach can transform UAVs from task-specific platforms into a continuously evolving communication network.
\end{abstract}


\section{Introduction}
\IEEEPARstart{R}{ecent} advances in artificial intelligence (AI) have significantly accelerated the evolution of unmanned aerial vehicle (UAV) networking. Learning-based approaches have achieved remarkable progress in various UAV-enabled networking tasks, including trajectory optimization, aerial base-station deployment, resource allocation, and multi-UAV coordination\cite{Abir23SDUAV6G,Song25TrustworthyLAE}. These advances demonstrate that UAVs are gradually evolving from conventional task-oriented platforms into intelligent network nodes. However, as low-altitude applications become increasingly diverse and networking environments become more dynamic and uncertain, optimizing individual UAV networking tasks is no longer sufficient to meet future requirements\cite{Song25TrustworthyLAE}. For example, providing reliable communication services in unknown regions requires not only trajectory planning or coverage optimization, but also autonomous discovery of communication demands, adaptive selection of UAV functional roles, and joint formation of access and backhaul networks. This challenge will become increasingly prominent in future sixth-generation (6G) networks and low-altitude economy scenarios, where communication infrastructures are expected to achieve autonomous operation, real-time adaptation, and reliable service provision under highly dynamic conditions. Therefore, future UAV networks need to move beyond predefined optimization toward intelligent networks with autonomous exploration, dynamic formation, and adaptive service.

Existing UAV networking research has achieved substantial progress in aerial base-station deployment, communication-aware trajectory design, mobile relaying, resource allocation, and multi-UAV coordination. Optimization-based approaches typically formulate UAV mobility and communication decisions as joint optimization problems to improve objectives such as throughput, coverage, fairness, and energy efficiency\cite{Zeng17EETrajectory,Wu18MultiUAV}. More recently, deep reinforcement learning (DRL) has enabled online decision-making in dynamic environments, while graph neural networks (GNNs) have provided scalable representations for capturing interactions among UAVs, users, and communication links\cite{Liu18DRLUAV,Wang21MARLUAV,Zhang23HGNN}. Despite these advances,  most existing approaches still rely on predefined network instances, where user sets, UAV configurations, and functional roles are determined in advance. Moreover, exploration, access provisioning, and backhaul construction are often optimized separately, limiting their ability to autonomously form and adapt heterogeneous UAV networks under evolving demands. Consequently, existing approaches have limited flexibility in adapting to dynamic environments and jointly optimizing multiple coupled networking objectives.

These limitations indicate that future UAV networks cannot be achieved solely through task-specific optimization methods designed for predefined network configurations. Instead, autonomous network formation requires an intelligent mechanism capable of continuously perceiving unknown environments, understanding evolving network states, coordinating exploration and service provisioning, and adapting network structures according to real-time feedback. Recent advances in large language models (LLMs) and LLM-based autonomous agents provide a promising pathway toward realizing such adaptive networking intelligence \cite{Wang24LLMAgent,Yang26LLMWComm}. With capabilities in contextual understanding, planning, memory, and tool utilization, LLM agents can integrate heterogeneous information, identify network bottlenecks, and coordinate subtasks such as exploration, UAV deployment, and topology optimization. Moreover, by continuously maintaining historical experiences and incorporating environmental feedback, the agent can progressively improve its understanding of network states and decision outcomes, optimize future exploration and deployment strategies, and enable heterogeneous UAV networks with autonomous construction, dynamic adaptation, and continuous evolution capabilities.

Based on this vision, this article investigates how LLM-assisted agentic AI can enable autonomous exploration and self-organizing heterogeneous UAV networks in low-altitude environments. We first review the evolution of UAV networking paradigms and analyze why existing approaches remain insufficient for autonomous network formation. We then discuss the unique capabilities of LLM-assisted agents and identify the key mechanisms required for autonomous exploration and heterogeneous UAV deployment. Building on these insights, we present an LLM-agent-assisted architecture for self-organizing heterogeneous UAV networks. Finally, we present a representative case study to illustrate the autonomous network-formation process, demonstrating how Agentic AI can support adaptive UAV organization in dynamic environments and laying the groundwork for practical, trustworthy, and continuously evolving low-altitude UAV networks.

\section{Evolution of UAV Networking Paradigms}

In this section, we review the evolution of UAV networking from optimization-driven designs to learning-based and emerging intelligent paradigms, and summarize representative methods and applications. We also discuss their key limitations in multi-task coordination, dynamic environments, and autonomous network formation, as summarized in Table ~\ref{tab1}.

\subsection{Evolution of UAV Networking Paradigms}

\subsubsection{Conventional Optimization-Driven UAV Networking}

Early UAV networking research mainly modeled UAVs as controllable aerial communication platforms and focused on optimizing their deployment locations, trajectories, relay topologies, and communication parameters. Representative studies addressed aerial base-station deployment, communication-aware trajectory planning, mobile relaying, and joint resource allocation. In these works, UAV mobility and wireless communication variables are typically formulated as coupled optimization problems to maximize throughput, improve coverage, enhance user fairness, or reduce energy consumption. Typical solution methods include alternating optimization, block coordinate descent, successive convex approximation, graph-based search, and other model-based optimization techniques. However, they usually assume that the user distribution, mission objective, UAV fleet size, and communication requirements are known in advance, and thus are mainly suitable for predefined network instances.

\subsubsection{Learning-Driven UAV Networking}

To improve the adaptability of UAV networks in dynamic environments, AI-enabled UAV networking has emerged as a promising direction for future development. Such networks allow UAVs to perceive surrounding environments, make autonomous decisions, and effectively collaborate, thereby supporting reliable communications in complex and time-varying scenarios. Early studies mainly employed deep reinforcement learning (DRL) for single-UAV or centrally controlled networking tasks, enabling UAVs to directly learn mobility and communication decisions from observed network states. For example, actor--critic-based DRL has been applied to jointly balance communication coverage, user fairness, connectivity, and UAV energy consumption, demonstrating the potential of learning-based control for online UAV networking \cite{Liu18DRLUAV}. This paradigm was subsequently extended to joint trajectory and communication optimization, where continuous-control methods such as deep deterministic policy gradient (DDPG) were adopted to jointly determine UAV mobility and communication-resource decisions.

As UAV networks evolved from single-platform operation toward cooperative multi-UAV systems, the learning paradigm further shifted from single-agent DRL to multi-agent reinforcement learning (MARL). Under this framework, each UAV acts as an independent decision-making agent and learns to coordinate its trajectory or communication behavior with other UAVs under a shared networking objective. For example, MARL has been applied to multi-UAV-assisted edge computing to jointly coordinate UAV trajectories and user-service decisions, while multi-UAV wireless data collection has been formulated as a decentralized partially observable decision-making problem, enabling UAV teams to cooperatively divide data-collection tasks and adapt their trajectories to different mission configurations \cite{Wang21MARLUAV,Bayerlein21MultiUAVDRL}.

More recently, graph-based learning has emerged as an important extension of MARL for UAV networking. Conventional neural-network policies typically represent observations of UAVs and users using fixed-dimensional vectors, making it difficult to efficiently capture pairwise interactions and adapt to dynamically changing network topologies. Graph neural networks (GNNs), in contrast, model UAVs, users, and their communication relationships as nodes and edges, enabling relational information to be aggregated through shared graph operations. Heterogeneous GNNs have therefore been combined with MARL to encode local UAV--user observations and facilitate information exchange among UAVs for cooperative trajectory design \cite{Zhang23HGNN}.

\subsubsection{LLM-Assisted UAV Networking}

 In recent years, advances in  LLMs, particularly in complex reasoning, knowledge transfer, and task planning, have begun to motivate their application to UAV networking and multi-agent coordination, although related research remains at an early stage. Existing studies mainly employ LLMs as high-level decision-enhancement modules within conventional optimization or learning frameworks. For example, LLMs have been integrated with multi-agent reinforcement learning to provide high-level decision knowledge and facilitate knowledge transfer, thereby improving the training efficiency and scalability of large-scale multi-hop UAV networking \cite{Xu26LLMUAV}. Other studies combine LLMs with game-theoretic optimization, where the model adaptively adjusts utility-function weights according to network states to assist UAV deployment and topology configuration \cite{Tang26AgenticDeployment}. In addition, LLMs have been incorporated into edge-assisted agentic architectures to support high-level reasoning, task coordination, and collaborative decision making in autonomous UAV swarms \cite{Nguyen26AgenticUAVSwarm}. These early efforts demonstrate the potential of LLMs to complement conventional optimization and learning strategies through higher-level semantic understanding, reasoning, and coordination, and to promote the evolution of UAV networking from task-specific decision making toward more adaptive and autonomous network intelligence. 

\subsection{Limitations of Existing Approaches}

In this subsection, we summarize these limitations from the perspectives of \emph{traditional optimization-based methods}, \emph{learning-based methods}, and \emph{LLM-assisted methods}, while their representative characteristics are summarized in Table~\ref{tab1}.

\textit{Traditional  methods:}
Traditional UAV networking approaches typically rely on predefined infrastructure, system models, and optimization objectives. UAV trajectories, deployments, relay topologies, and communication resources are optimized under known conditions, but their dependence on prior environmental knowledge limits adaptability in dynamic low-altitude scenarios. When user demands or network topology changes, these methods often require reformulating the optimization problem, making autonomous exploration, network expansion, and real-time reconfiguration difficult.

\textit{Learning-based methods:}
Learning-driven approaches, including DRL, MARL, and graph-based policies, improve online decision making and adaptability in UAV networks. However, most existing studies assume predefined agents, user information, UAV resources, and functional roles, focusing on optimizing the behavior of an existing network rather than determining when and how to expand it. Moreover, task-specific designs with fixed state, action, and reward spaces limit their ability to jointly handle unknown environment exploration, heterogeneous UAV deployment, access provisioning, and backhaul formation as an integrated network-formation process.

\textit{LLM-assisted methods:}
Recent LLM-assisted UAV networking approaches provide a new pathway for integrating high-level knowledge into optimization and learning frameworks. However, existing studies mainly use LLMs as auxiliary modules, while networking tasks, agent roles, and action spaces remain predefined. Such designs limit the ability of LLMs to autonomously reason about network reconfiguration and lack a persistent perception--reasoning--action--feedback loop for adapting topology and resources to evolving communication demands.

\begin{table*}[!t]
\centering
\caption{Comparison of representative UAV networking paradigms.}
\label{tab1}

\scriptsize
\setlength{\tabcolsep}{4pt}
\renewcommand{\arraystretch}{1.35}

\begin{tabularx}{\textwidth}{
C{0.12\textwidth}
>{\centering\arraybackslash}p{0.18\textwidth}
>{\raggedright\arraybackslash}X
>{\raggedright\arraybackslash}X
>{\raggedright\arraybackslash}X
}

\toprule

\textbf{Paradigm} &
\textbf{Representative Tasks} &
\textbf{Typical Methods} &
\textbf{Advantages} &
\textbf{Limitations}
\\

\midrule

\rowcolor{RowBlue}
\textbf{Traditional}
&
Trajectory design, UAV deployment, resource allocation, relay and backhaul formation
&
Alternating optimization, BCD, SCA, and model-/game-based optimization
\cite{Zeng17EETrajectory,Wu18MultiUAV,Challita17Backhaul}
&
$\bullet$ Physically interpretable and engineering-friendly. \newline
$\bullet$ Explicitly incorporates mobility, communication, and energy constraints. \newline
$\bullet$ Stable under well-defined network settings.
&
$\bullet$ Requires substantial prior network and environmental knowledge. \newline
$\bullet$ User sets, UAV numbers, functional roles, and objectives are mostly predefined. \newline
$\bullet$ Limited adaptability to unknown demands and dynamic network expansion.
\\

\midrule

\rowcolor{RowPeach}
\textbf{Learning-Based}
&
Online UAV control, cooperative trajectory planning, service coordination, and topology-aware multi-UAV control
&
DRL for online UAV control \cite{Liu18DRLUAV};
MARL for distributed multi-UAV coordination \cite{Wang21MARLUAV};
GNN-enhanced learning for topology-aware decisions \cite{Zhang23HGNN}
&
$\bullet$ Supports online adaptation to time-varying network states. \newline
$\bullet$ Enables distributed multi-UAV cooperation. \newline
$\bullet$ Graph representations improve scalability to changing network topologies.
&
$\bullet$ Agent sets, state--action spaces, roles, and reward functions remain predefined. \newline
$\bullet$ Usually addresses a specific networking task. \newline
$\bullet$ Limited capability for unknown-demand discovery and on-demand heterogeneous network formation.
\\

\midrule

\rowcolor{RowLav}
\textbf{LLM-Assisted}
&
Multi-hop networking, UAV deployment and topology optimization, and autonomous swarm coordination
&
LLM-to-MARL knowledge transfer \cite{Xu26LLMUAV};
LLM-enhanced game-theoretic deployment \cite{Tang26AgenticDeployment};
edge-assisted LLM agent coordination \cite{Nguyen26AgenticUAVSwarm}
&
$\bullet$ Introduces high-level knowledge and contextual reasoning. \newline
$\bullet$ Complements existing optimization and learning frameworks. \newline
$\bullet$ Provides a pathway toward more adaptive network coordination.
&
$\bullet$ LLMs mainly serve as auxiliary modules for predefined tasks. \newline
$\bullet$ Contextual reasoning, memory, and long-horizon coordination remain underused. \newline
$\bullet$ Persistent feedback-driven network evolution is still largely unexplored.
\\

\bottomrule

\end{tabularx}
\end{table*}

\section{Vision of Self-Organizing UAV Networks}
\label{sec:vision}

\begin{figure*}[htbp]
    \centering
    \includegraphics[width=0.96\textwidth]{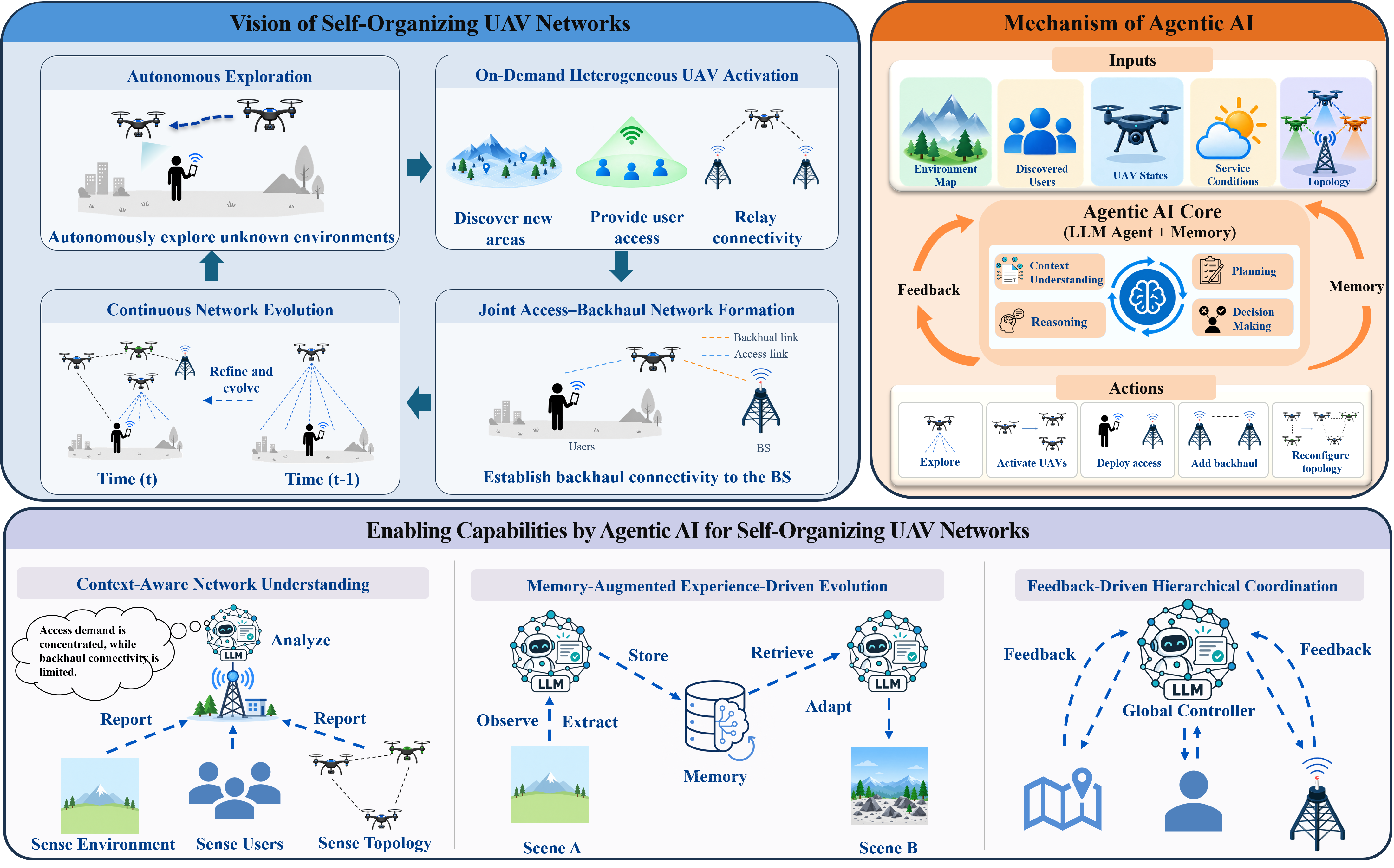} 
    \caption{Agentic AI empowered autonomous networking: vision, mechanism, and enabling capabilities.}
    \label{fig1}
\end{figure*}

The limitations discussed above suggest that future UAV networks should move beyond predefined missions and fixed configurations. In unknown and dynamic low-altitude environments, user demands emerge progressively, while network scale, functional resources, and connectivity cannot be fully determined in advance. We therefore envision a self-organizing UAV network that explores the environment, activates resources on demand, forms end-to-end communication infrastructure, and continuously adapts its organization during operation. As illustrated in Fig.~\ref{fig1}, such a network should possess the following key capabilities.

\subsubsection{\textbf{Autonomous Exploration}}

Future UAV networks should be capable of actively exploring initially unknown environments rather than relying on complete prior knowledge of user locations or communication demands. Exploration UAVs can progressively collect environmental information and discover previously unknown users, allowing networking decisions to be driven by actually observed demands. This capability transforms environment exploration from an offline prerequisite into an integral part of network operation and provides the information basis for subsequent network formation.

\subsubsection{\textbf{On-Demand Heterogeneous UAV Activation}}

The network scale should not remain fixed throughout a mission. Instead of deploying the entire UAV fleet in advance, aerial resources can be activated progressively as communication demands emerge. Newly discovered users may trigger additional coverage resources, while insufficient connectivity to the terrestrial infrastructure may require additional backhaul UAVs. Consequently, the number and functions of active UAVs become adaptive to the evolving network state, enabling more flexible and resource-efficient network deployment.

\subsubsection{\textbf{Joint Access--Backhaul Network Formation}}

Discovering or locally covering users alone does not guarantee effective communication service. A self-organizing UAV network should jointly construct user access and backhaul connectivity so that discovered users can establish valid end-to-end communication paths to the terrestrial network. Access and backhaul formation should therefore be treated as coupled components of the same network-formation process rather than as independent optimization tasks. As communication demands are progressively revealed, the aerial topology can expand and reorganize to maintain end-to-end service availability.



\subsubsection{\textbf{Continuous Network Evolution}}

Self-organization should not terminate once an initial network has been formed. As additional users are discovered, service conditions change, or previous deployment decisions prove ineffective, the network should evaluate its current organization and reconfigure itself accordingly. More importantly, useful outcomes from previous decisions can be retained and reused when similar conditions arise, allowing operational experience to progressively influence future network formation. 

\section{Enabling Self-Organizing UAV Networks with Agentic AI}
\label{sec:agentic_ai}

The challenges identified above require UAV networks to move beyond predefined configurations and isolated task optimization. As communication demands are progressively revealed, the network must continuously interpret new information, coordinate local and global decisions, and adapt its organization based on operational feedback. Agentic AI is particularly well suited to this setting, as its closed-loop perception, reasoning, planning, action, feedback, and memory provide the intelligence needed for persistent network understanding and adaptation. We next introduce the basic mechanism of agentic AI and discuss how it enables autonomous exploration, on-demand network formation, and continuous network evolution.

\subsection{Mechanism of Agentic AI}
The core of agentic AI is a persistent, goal-driven loop of perception, reasoning, planning, action, and feedback. The agent integrates environmental observations, system states, objectives, constraints, and historical experience into a unified context. Reasoning identifies current needs, planning converts them into executable actions, and the resulting outcomes are fed back to support subsequent decisions. Useful interaction history can be retained in memory. Unlike conventional LLM-assisted methods used for isolated functions, an LLM-based agent remains continuously involved through this memory-supported closed loop, providing a foundation for autonomous network formation and continuous evolution.

\subsection{Enabling Capabilities by Agentic AI for Self-Organizing UAV Networks}

Building on this agentic mechanism, self-organizing UAV networks can be endowed with three key capabilities: context-aware network understanding, memory-augmented experience-driven evolution, and feedback-driven hierarchical coordination across different decision timescales. Together, these capabilities enable UAV networks to autonomously interpret evolving network conditions, learn from operational experience, and coordinate global adaptation with fast local control. These capabilities are discussed in the following.

\subsubsection{\textbf{Context-Aware Network Understanding}}
Context-aware reasoning enables the agent to interpret the UAV network as an evolving system by integrating environmental information, exploration progress, discovered users, UAV states, service conditions, and access--backhaul relationships into a structured network context. Based on this context, the agent can identify whether the current limitation arises from insufficient exploration, inadequate access coverage, constrained backhaul connectivity, or inefficient network organization. This capability supports autonomous exploration, on-demand UAV activation, and joint network formation. For example, newly discovered users may indicate insufficient access resources, while weak connectivity to the terrestrial base station may reveal a backhaul shortage. Therefore, the agent can dynamically adapt UAV deployment, access--backhaul relationships, and operational priorities according to observed communication demands, enabling autonomous network evolution beyond predefined configurations.

\subsubsection{\textbf{Memory-Augmented Experience-Driven Evolution}}

Memory enables agentic AI to retain task context, interaction history,
and decision outcomes across decision cycles. Short-term memory supports ongoing operation, while long-term memory stores reusable experience summarized from previous network states and outcomes. At each decision cycle, relevant historical experience can be retrieved as additional context for reasoning, allowing the agent to reuse effective configurations and avoid repeated exploration or redundant resource activation. By associating network configurations with their observed effects on coverage, connectivity, and service quality, memory enables experience-driven adaptation and extends self-organization toward continuous network evolution.

\subsubsection{\textbf{Feedback-Driven and Hierarchical Coordination}}

Feedback-driven interaction enables the agent to evaluate the outcomes of network decisions and refine future actions. After each adjustment, changes in coverage, connectivity, and service performance are fed back to the reasoning process, allowing further reconfiguration when the expected improvement is not achieved. This continuous verification loop transforms network operation from one-shot configuration into adaptive evolution. However, such adaptation involves heterogeneous decision timescales: mobility and local coordination require fast responses, whereas topology assessment and infrastructure reconfiguration require global reasoning. A hierarchical architecture can therefore assign rapid local control to lightweight agents and global adaptation to a higher-level agent, achieving both real-time responsiveness and network evolution.

\section{Case Study: Autonomous Network Formation in an Unknown Environment}
\label{sec:case_study}

\subsection{Proposed System}

\begin{figure*}[htbp]
    \centering
    \includegraphics[width=0.98\textwidth]{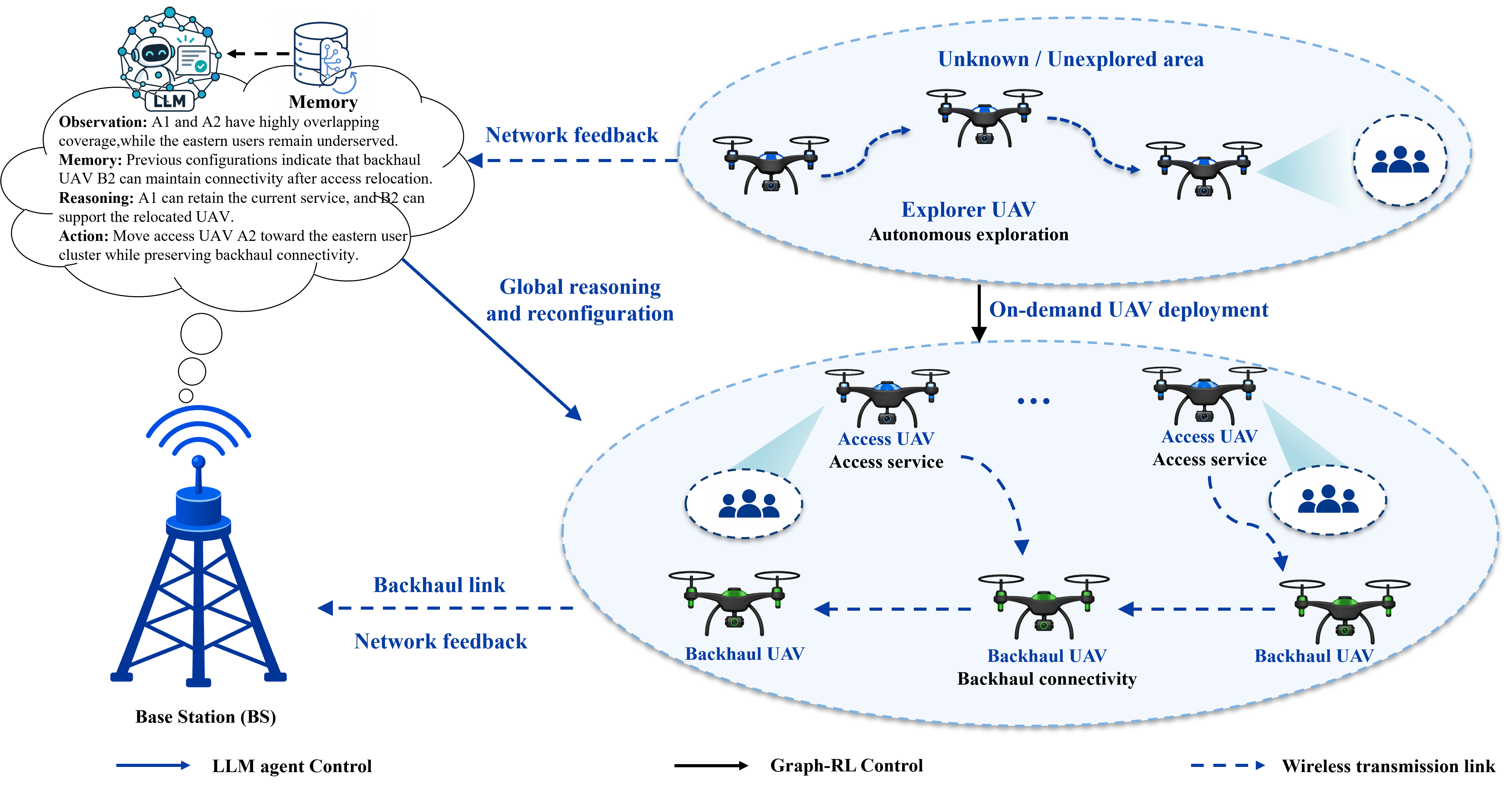} 
    \caption{Proposed self-organizing UAV network. }
    \label{fig2}
\end{figure*}

Building on the above vision, we develop a representative implementation for self-organizing UAV network formation in unknown low-altitude environments. As shown in Fig.~\ref{fig2}, the system consists of a heterogeneous UAV swarm, onboard Graph-RL agents, and a base-station LLM agent. Through hierarchical coordination, it progressively discovers demands, activates UAV resources on demand, and establishes end-to-end access and backhaul connectivity.
\subsubsection{\textbf{Heterogeneous UAV Swarm}}

The UAV swarm consists of three functionally heterogeneous types of UAVs: exploration UAVs, access UAVs, and backhaul UAVs. Exploration UAVs are responsible for actively exploring unknown regions, discovering users, and collecting locally available network information. In addition to mobility decisions, they determine when additional exploration, access, or backhaul UAVs should be activated according to the communication demands and network conditions revealed during exploration.
Access UAVs are deployed to provide wireless access to discovered users, while backhaul UAVs establish relay links between the aerial access network and the terrestrial base station. Together, these two types of UAVs construct the end-to-end service path from users to the terrestrial network. Rather than being fully deployed in advance, access and backhaul UAVs are progressively
activated as new communication demands emerge. After deployment, access and backhaul UAVs mainly maintain the formed communication infrastructure. Their configurations can subsequently be adjusted by the base-station LLM agent when the current access or backhaul topology becomes inefficient.

\subsubsection{\textbf{Onboard Graph-RL Agents}}

Lightweight Graph-RL agents are deployed on the exploration UAVs to provide fast and distributed decision making during network formation. Each agent determines the exploration movement of its UAV and decides when additional exploration, access, or backhaul UAVs should be activated according to the locally observed communication demands and network conditions. As users are progressively discovered and new UAVs are activated, both the
network scale and communication topology continuously change. Graph-based representations naturally capture these dynamic relationships, allowing each agent to combine its local observation with information exchanged among reachable neighboring UAVs. To support distributed decisions under varying network scales and topologies, the exploration UAVs share a common policy,
while graph message passing enables local information exchange and coordination. The agents are trained following a centralized-training-and-decentralized-execution paradigm. Global network information is available during training to evaluate cooperative decisions, whereas each exploration UAV relies only on its local observations and graph messages during execution. 

\subsubsection{\textbf{Base-Station LLM Agent}}
The LLM agent is deployed at the terrestrial base station and operates at a lower decision frequency than the graph policy. As exploration proceeds, information collected by the UAV swarm is progressively aggregated into a structured global network context, including exploration progress, discovered and unserved users, active UAV roles and positions, available UAV resources, access relationships, backhaul connectivity, and current service conditions. Based on this context, the LLM agent evaluates whether the network formed by the graph policy remains appropriate as additional information becomes available. Rather than controlling routine exploration movements, it focuses on network-level reconfiguration. For example, the agent can identify redundant coverage, inefficient infrastructure placement, or insufficient backhaul connectivity and accordingly adjust existing access or backhaul UAVs. The LLM agent further exploits feedback and historical experience to support network evolution. After a network adjustment, the resulting changes in coverage, connectivity, and service performance are associated with the corresponding network condition and decision. Useful experiences are retained as high-level operational knowledge and retrieved to guide future decisions under similar network conditions.

\subsection{Simulation Setup}
\begin{table*}[t]
\centering
\caption{Planning performance over 42 paired test scenarios.}
\label{tab:planning_comparison}
\begin{tabular}{lccccc}
\toprule
Method
& Discovery (\%) $\uparrow$
& Coverage (\%) $\uparrow$
& Throughput (Mbps) $\uparrow$
& Cumulative throughput (Mbps-slot) $\uparrow$
& Average  UAVs $\downarrow$ \\
\midrule
CCGD
& $65.34$
& $41.49$
& $13.40$
& $1040.72$
& $\mathbf{6.17}$ \\

GNN
& $90.50$
& $61.91$
& $19.61$
& $1441.96$
& $8.67$ \\

Proposed
& $\mathbf{93.53}$
& $\mathbf{70.66}$
& $\mathbf{22.04}$
& $\mathbf{1536.84}$
& $8.76$ \\
\midrule

OAGP$^{\dagger}$
& $98.51$
& $75.96$
& $22.95$
& $1602.43$
& $8.29$ \\
\bottomrule
\end{tabular}

\vspace{1mm}
\footnotesize{$^{\dagger}$Uses hidden ground-truth user locations and is
not a certified global optimum. Bold denotes the best deployable method.
}
\end{table*}
\begin{table*}[t]
\centering
\caption{Ablation study over the same 42 paired test scenarios.}
\label{tab:memory_ablation}
\begin{tabular}{lccccc}
\toprule
Method
& Discovery (\%) $\uparrow$
& Coverage (\%) $\uparrow$
& Throughput (Mbps) $\uparrow$
& Cumulative throughput (Mbps-slot) $\uparrow$
& Average UAVs $\downarrow$ \\
\midrule
GNN
& $90.50$
& $61.91$
& $19.61$
& $1441.96$
& $\mathbf{8.67}$ \\

Proposed w/o memory
& $92.75$
& $64.87$
& $20.81$
& $1501.53$
& $8.76$ \\

Proposed
& $\mathbf{93.53}$
& $\mathbf{70.66}$
& $\mathbf{22.04}$
& $\mathbf{1536.84}$
& $8.76$ \\
\bottomrule
\end{tabular}

\vspace{1mm}

\end{table*}
We evaluate the proposed framework in an $800~\mathrm{m} \times 800~\mathrm{m}$ low-altitude region over 80 time slots of $5~\mathrm{s}$ each. User and terrestrial base-station locations are fixed within each scenario and randomly generated across scenarios. The UAV fleet contains up to three exploration, four access, and five backhaul UAVs. Initially, only one exploration UAV is active, while the remaining exploration, access, and backhaul UAVs are activated on demand as the network evolves. The graph policy is trained with 9--15 users and tested with 7--20 users. Three common random seeds are used for each user population, yielding 42 paired test scenarios. The graph controller follows centralized training with decentralized execution (CTDE) and is trained for $1.5 \times 10^6$ environment steps using Adam with a learning rate of $2 \times 10^{-4}$. Qwen3-14B serves as the base-station LLM agent and is invoked every ten slots for global planning and infrastructure reconfiguration, while onboard Graph-RL agents make per-slot decisions. We compare against three baselines: (i) \emph{CCGD}, a grid-based greedy planner that explores unvisited regions and activates  access and backhaul UAVs on demand; (ii) \emph{GNN}, the distributed graph controller without LLM assistance; and (iii) \emph{OAGP}, a full-information planner that uses ground-truth user locations for access and backhaul planning. Since such information is unavailable in practice, OAGP serves only as a noncausal reference. We evaluate five metrics: \emph{Discovery}, the fraction of discovered users; \emph{Coverage}, the fraction with both wireless access and a valid backhaul path; \emph{Throughput}, the final aggregate end-to-end throughput; \emph{Cumulative throughput}, the throughput accumulated during network formation; and \emph{Average UAVs}, the average number of UAVs.

\begin{figure*}[!t]
\centering

\begin{minipage}[t]{0.31\textwidth}
    \centering
    \includegraphics[width=\linewidth]{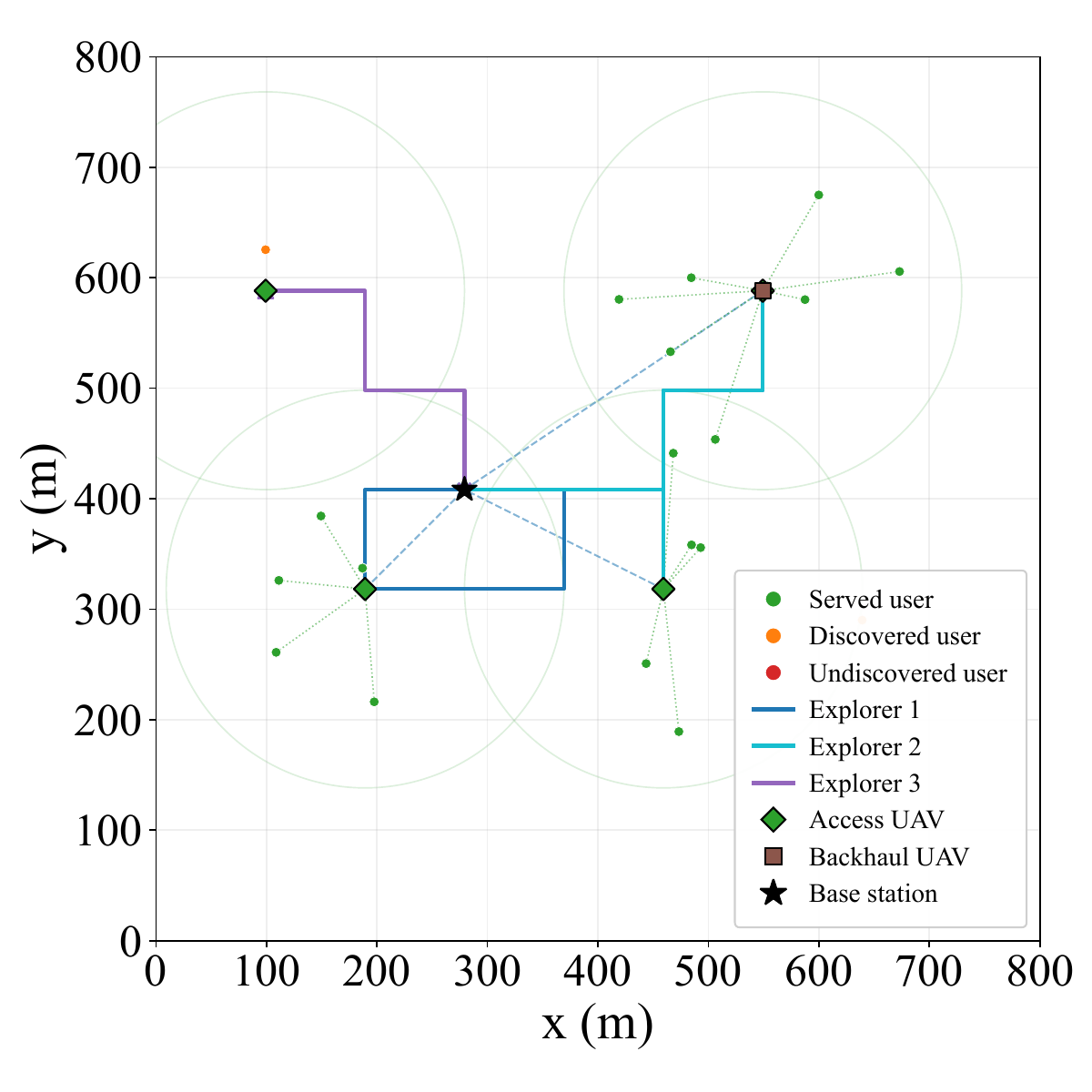}

    \vspace{0.5mm}
    {(a) OAGP}

    \vspace{0.5mm}
    {\scriptsize
    \textit{Cov.} 89.47\%
    \;/\;
    \textit{Thr.} 31.96 Mbps
    \;/\;
    \textit{Cum.} 2249.94 Mbps-slot
    }
\end{minipage}
\hfill
\begin{minipage}[t]{0.31\textwidth}
    \centering
    \includegraphics[width=\linewidth]{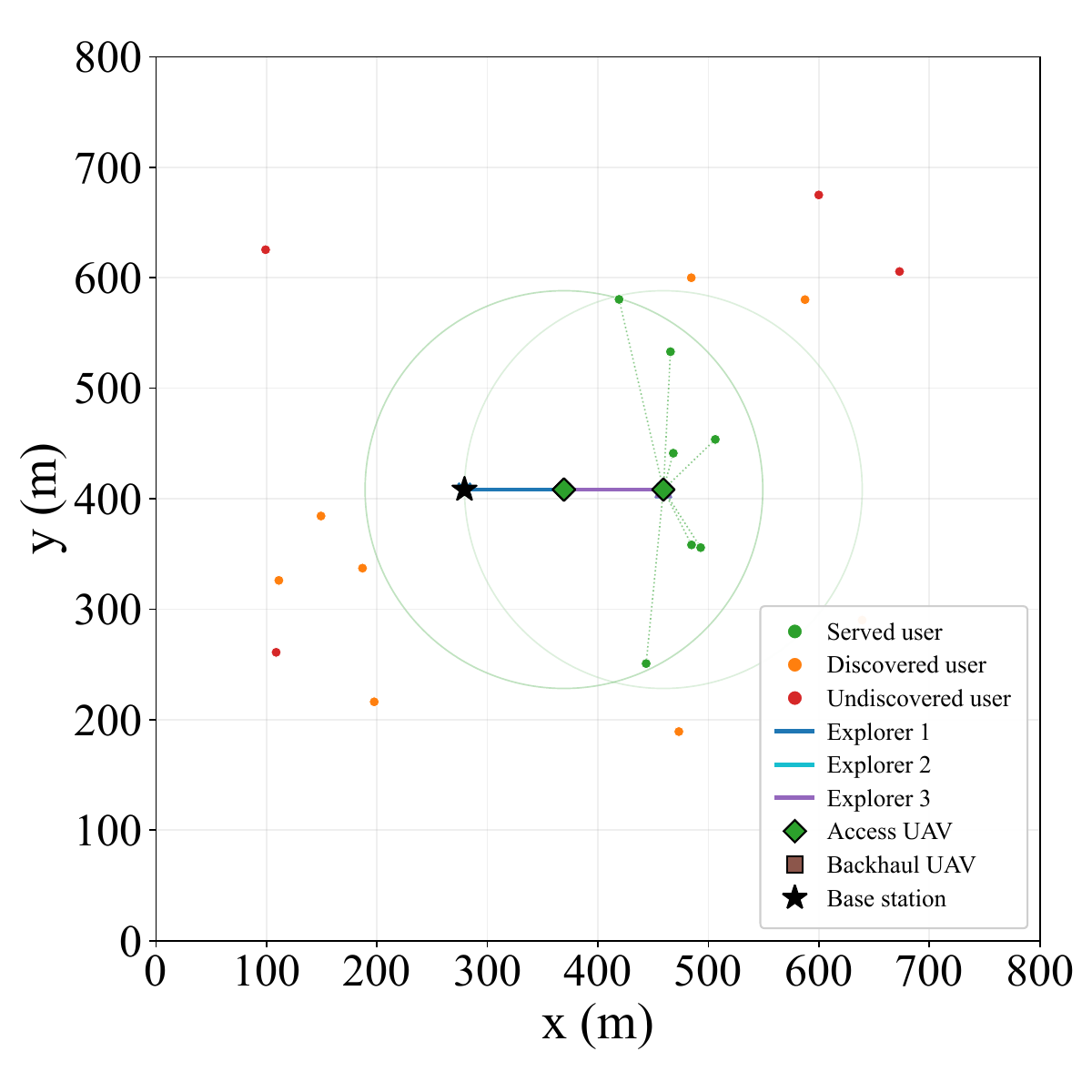}

    \vspace{0.5mm}
    {(b) CCGD}

    \vspace{0.5mm}
    {\scriptsize
    \textit{Cov.} 36.84\%
    \;/\;
    \textit{Thr.} 12.20 Mbps
    \;/\;
    \textit{Cum.} 956.73 Mbps-slot
    }
\end{minipage}
\hfill
\begin{minipage}[t]{0.31\textwidth}
    \centering
    \includegraphics[width=\linewidth]{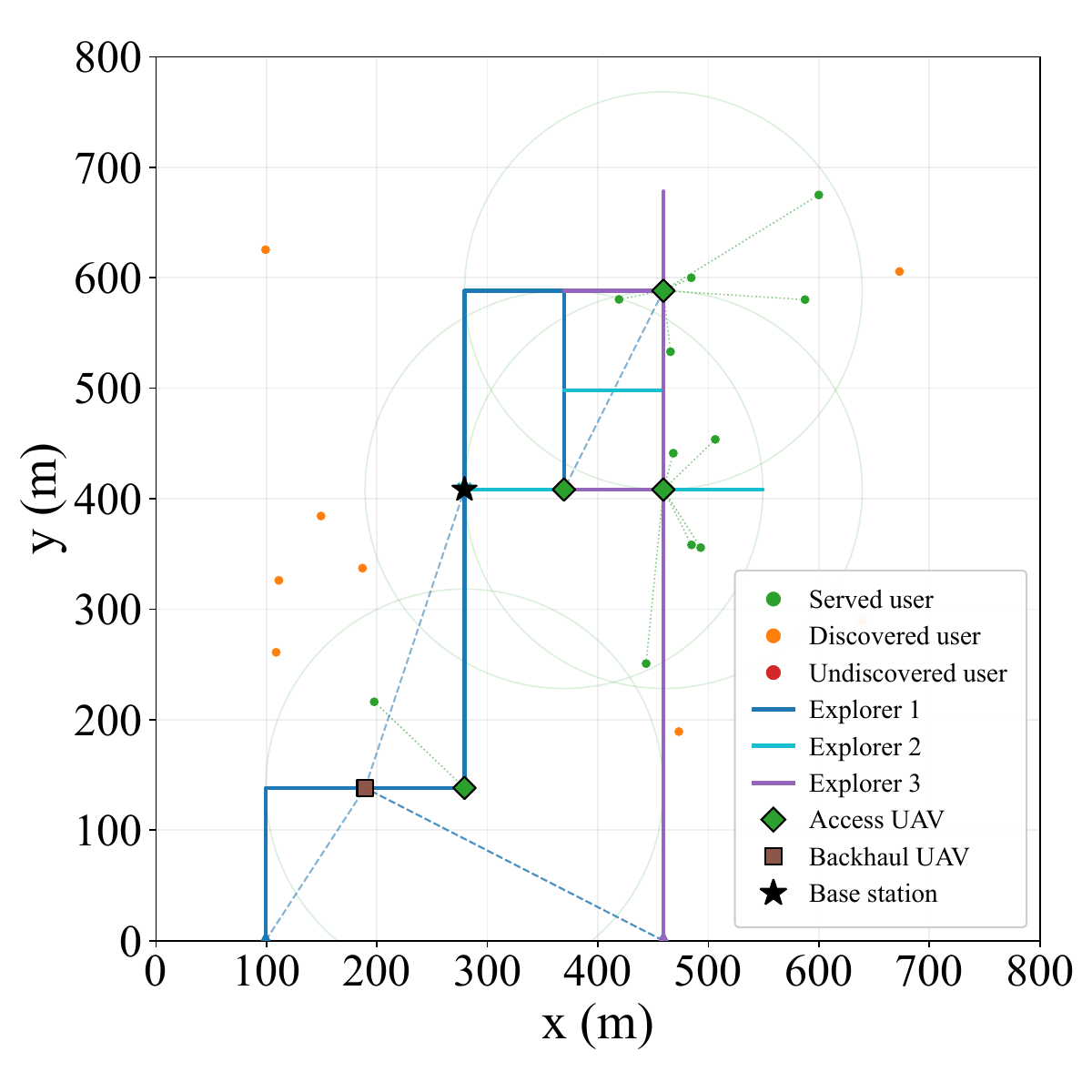}

    \vspace{0.5mm}
    {(c) GNN}

    \vspace{0.5mm}
    {\scriptsize
    \textit{Cov.} 57.89\%
    \;/\;
    \textit{Thr.} 22.93 Mbps
    \;/\;
    \textit{Cum.} 1745.40 Mbps-slot
    }
\end{minipage}

\vspace{3mm}

\begin{minipage}[t]{0.31\textwidth}
    \centering
    \includegraphics[width=\linewidth]{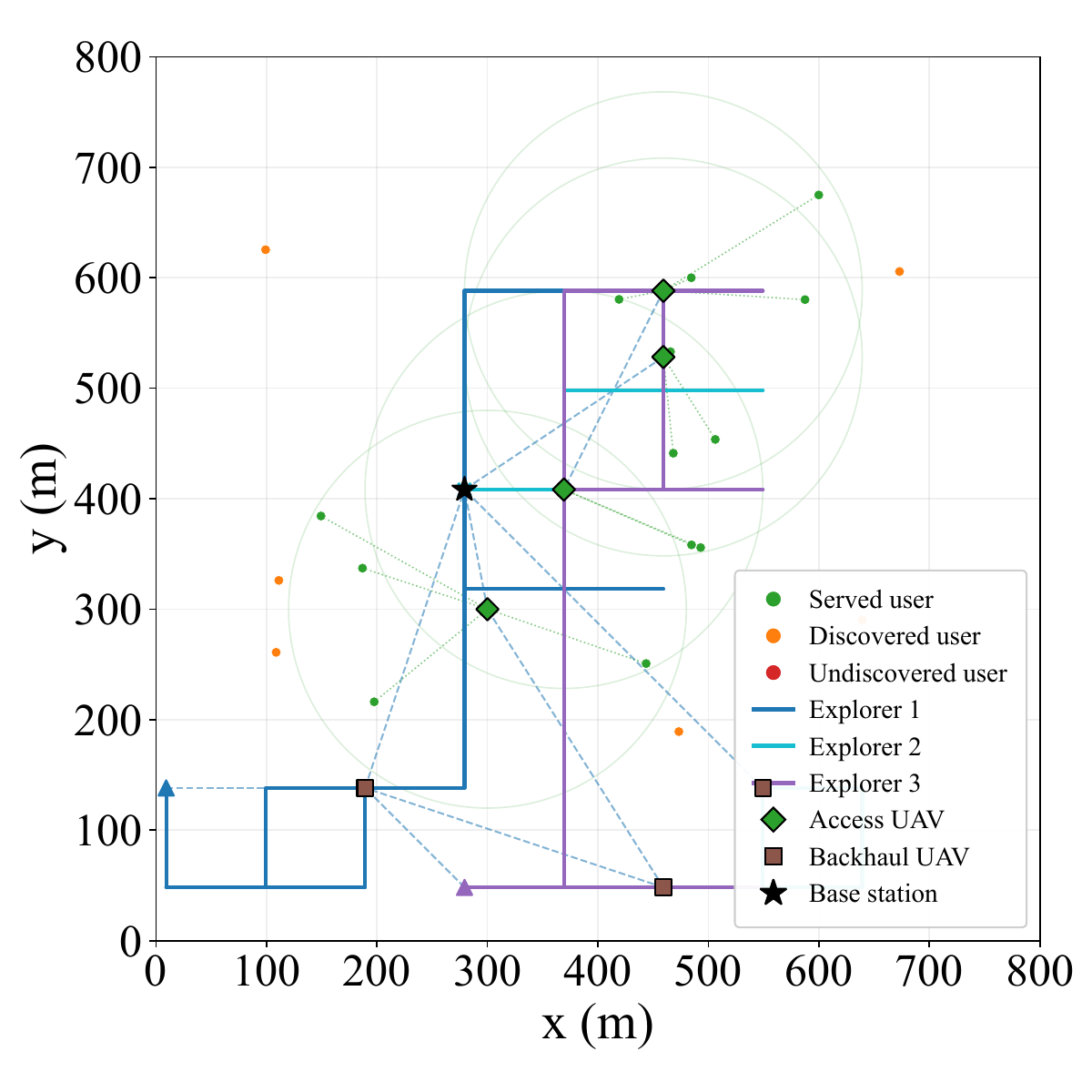}

    \vspace{0.5mm}
    {(d) Proposed w/o memory}

    \vspace{0.5mm}
    {\scriptsize
    \textit{Cov.} 68.42\%
    \;/\;
    \textit{Thr.} 31.50 Mbps
    \;/\;
    \textit{Cum.} 1871.96 Mbps-slot
    }
\end{minipage}
\hspace{0.035\textwidth}
\begin{minipage}[t]{0.31\textwidth}
    \centering
    \includegraphics[width=\linewidth]{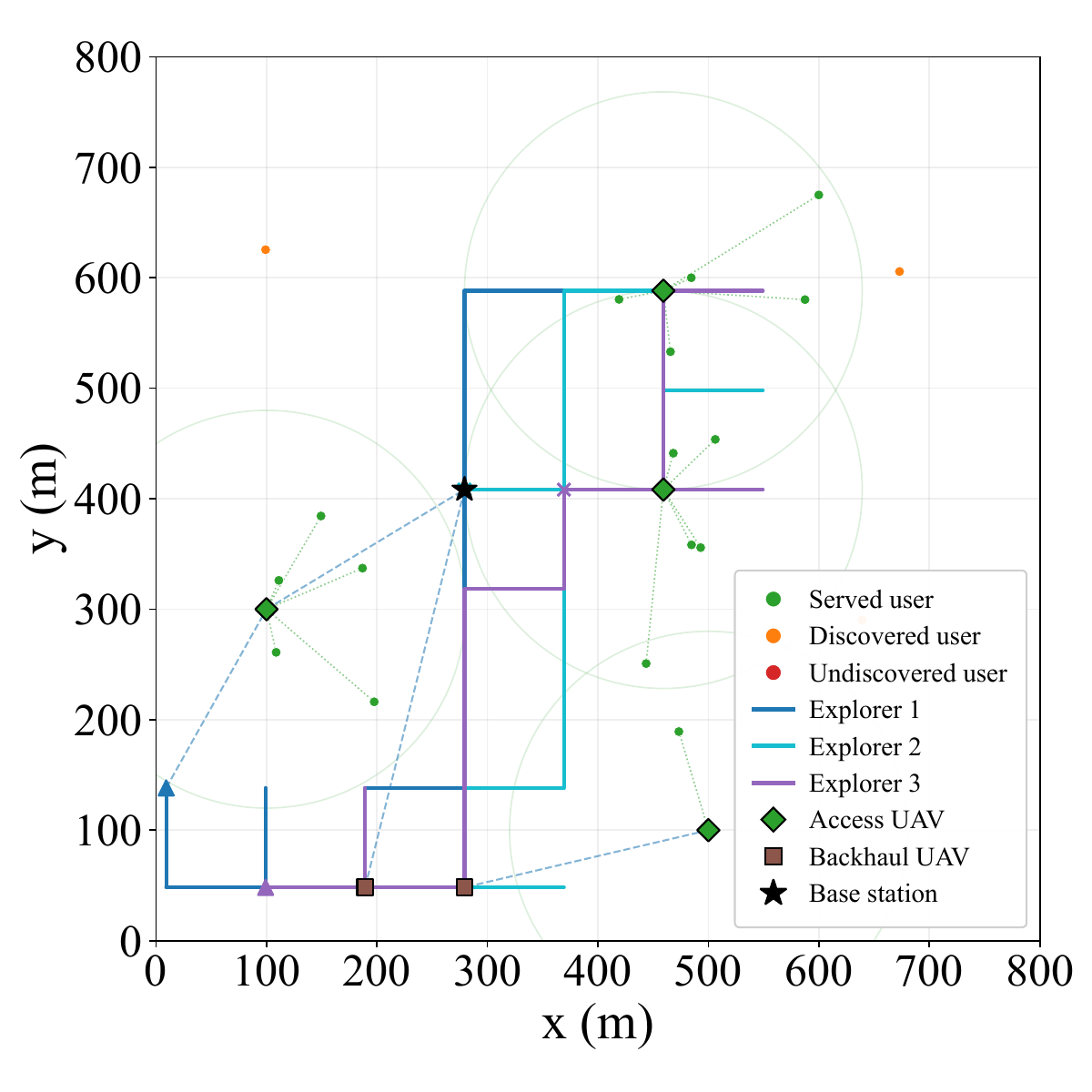}

    \vspace{0.5mm}
    {(e) Proposed}

    \vspace{0.5mm}
    {\scriptsize
    \textit{Cov.} \textbf{84.21\%}
    \;/\;
    \textit{Thr.} \textbf{33.74 Mbps}
    \;/\;
    \textit{Cum.} \textbf{2403.99 Mbps-slot}
    }
\end{minipage}

\caption{Representative network-formation results for five planning methods.
}
\label{fig:representative_trajectories}
\end{figure*}

\subsubsection{\textbf{Overall Performance}}

As shown in Table~\ref{tab:planning_comparison}, the proposed framework achieves the best performance among all deployable methods in terms of discovery, coverage, and throughput. Compared with CCGD, it improves discovery and coverage by 28.19\% and 29.17\%, respectively, while increasing throughput by 64.5\% and cumulative throughput by 47.7\%. Although CCGD activates fewer UAVs on average, its lower service performance indicates that geometry-based deployment is insufficient for jointly handling exploration, access provisioning, and backhaul formation under unknown demand. Compared with the standalone GNN, the proposed method improves discovery by 3.03\%, coverage by 8.75\%, throughput by 12.4\%, and cumulative throughput by 6.6\%, with only a marginal increase in average UAV usage from 8.67 to 8.76. The larger improvement in coverage than discovery indicates that the LLM mainly enhances the conversion of discovered users into effective access--backhaul service rather than simply increasing exploration. The oracle-based global planner achieves the highest performance due to its access to complete user locations. Nevertheless, without privileged information, the proposed method achieves approximately 94.9\% of its discovery, 93.0\% of its coverage, 96.0\% of its throughput, and 95.9\% of its cumulative throughput, demonstrating effective autonomous network
formation under partial observability.

\subsubsection{\textbf{Ablation Study}}

Table~\ref{tab:memory_ablation} further separates the effects of LLM-based global reasoning and feedback memory. Introducing state-dependent LLM planning without memory improves discovery by 2.25\%, coverage by 2.96\%, throughput by 6.1\%, and cumulative throughput by 4.1\% over the standalone GNN, while increasing the average number of active UAVs by only 0.09. Adding feedback memory further improves coverage by 5.79\% and throughput by 5.9\%, while discovery and cumulative throughput increase by 0.78\% and 2.35\%, respectively. Importantly, the average number of active UAVs remains unchanged at 8.76. This result shows that the benefit of memory does not arise from deploying additional resources; instead, it enables the LLM agent to make more effective use of the existing UAV fleet by associating previous reconfiguration decisions with their observed service outcomes.

\subsubsection{\textbf{Representative Network-Formation Cases}}
Fig.~\ref{fig:representative_trajectories} presents a representative
19-user scenario and highlights clear differences in network formation. With access to all user locations, OAGP directly places access UAVs around separated user clusters as a full-information reference. In contrast, CCGD forms a highly concentrated network around the base station and
the central-right region, leaving several peripheral users undiscovered or unserved. GNN explores a larger area, but its coverage and backhaul resources remain concentrated around the central and upper-right
regions, indicating that distributed local decisions alone may not achieve efficient global network organization. Introducing LLM-based planning produces a broader access--backhaul structure,
with the Proposed w/o memory method achieving 68.42\% coverage and 31.50~Mbps throughput. With feedback memory, the complete Proposed method further spreads access UAVs toward the separated left, upper-right, and lower-right user regions. This behavior can benefit from summarized experience retrieved from similar previous network conditions, allowing the agent to avoid repeatedly concentrating resources in already well-served areas and to favor broader coverage when appropriate. Although OAGP achieves slightly higher coverage using hidden user locations, the proposed method attains higher throughput in this case, demonstrating the benefit of memory-assisted global reconfiguration under partial observability.

\section{Conclusion}

In this article, we have investigated how agentic AI can support the autonomous evolution of UAV networks in low-altitude environments. We first reviewed the progression from optimization-based to learning-driven and LLM-assisted approaches, and highlighted their limitations in unknown and dynamically evolving networks. We then discussed key agentic capabilities, including context-aware reasoning, memory-augmented adaptation, and hierarchical coordination. Building on these insights, we presented an LLM-agent-enabled self-organizing UAV architecture and a representative case study showing progressive user discovery, on-demand resource activation, and end-to-end network formation. The results indicate that combining Graph-RL with LLM-based global coordination improves coverage and throughput under varying network scales. Overall, agentic AI shows strong potential for autonomous network formation and experience-driven adaptation, paving the way toward practical and continuously evolving low-altitude networks.

%
\IEEEpeerreviewmaketitle

\bibliographystyle{IEEEtran}
\bibliography{references}

\end{document}